\documentclass[]{spie}  

\usepackage{amssymb}
\usepackage{pifont}

\usepackage{amsmath,amsfonts,amssymb}
\usepackage{subcaption}
\usepackage{graphicx}
\usepackage{cite} 
\usepackage{times}
\usepackage{epsfig}
\usepackage{amsmath}
\usepackage{nccmath}
\usepackage{amssymb}
\usepackage{mwe}
\usepackage{acro}
\usepackage{amssymb}
\usepackage{xcolor,colortbl}
\usepackage{tabularx}
\usepackage{relsize}
\usepackage{pifont}
\usepackage{booktabs} 
\usepackage{multirow}
\usepackage{multicol}
\usepackage{adjustbox}
\usepackage{float}
\usepackage{graphicx}
\usepackage{makecell}
\usepackage{tabu}
\usepackage[colorlinks=true, allcolors=blue]{hyperref}
\usepackage[capitalize]{cleveref}

\title{Cross-modality Attention-based Multimodal Fusion for Non-small Cell Lung Cancer (NSCLC) Patient Survival Prediction}

\author[a,b]{Ruining Deng}
\author[a]{Nazim Shaikh}
\author[a]{Gareth Shannon}
\author[a]{Yao Nie}

\affil[a]{Roche Diagnostics Solutions, Santa Clara, CA, 95050, USA}
\affil[b]{Vanderbilt University, Nashville, TN, 37235, USA}

\begin{document} 
\maketitle

\begin{abstract}
Cancer prognosis and survival outcome predictions are crucial for therapeutic response estimation and for stratifying patients into various treatment groups. Medical domains concerned with cancer prognosis are abundant with multiple modalities, including pathological image data and non-image data such as genomic information. To date, multimodal learning has shown potential to enhance clinical prediction model performance by extracting and aggregating information from different modalities of the same subject. This approach could outperform single modality learning, thus improving computer-aided diagnosis and prognosis in numerous medical applications. In this work, we propose a cross-modality attention-based multimodal fusion pipeline designed to integrate modality-specific knowledge for patient survival prediction in non-small cell lung cancer (NSCLC). Instead of merely concatenating or summing up the features from different modalities, our method gauges the importance of each modality for feature fusion with cross-modality relationship when infusing the multimodal features. Compared with single modality, which achieved c-index of 0.5772 and 0.5885 using solely tissue image data or RNA-seq data, respectively, the proposed fusion approach achieved c-index 0.6587 in our experiment, showcasing the capability of assimilating modality-specific knowledge from varied modalities.

\end{abstract}



  


\section{Description of purpose}
\label{sec:intro}  
Cancer prognosis and survival outcome predictions are crucial for therapeutic response forecasts and stratifying patients into distinct treatment groups. Previous works~\cite{mobadersany2018predicting,cheerla2019deep,Chen_2021_ICCV} have demonstrated that incorporating various data modalities into survival prediction models can bolster their predictive capabilities, thereby benefiting both clinical research and practice. However, these methods simply concatenate or sum up the features from different modalities, neglecting a deeper understanding of inter-modality interactions during the fusion process. The latest attention-based methods~\cite{ilse2018attention,jaume2023modeling} show promising fusion performance by discerning the relationships between different modalities. In this research, our aim was to predict the survival outcomes of patients with NSCLC using a blend of histopathology and genomics. We employed a cross-modality attention-based multimodal fusion (CM-MMF) approach, integrating image and RNA-seq modalities to achieve superior patient survival predictions. This showcases the potential of assimilating modality-specific knowledge from varied sources. The attention scores derived from the fusion layer highlight the significance of each modality during fusion for clinical diagnosis.

\begin{figure}[h]
\begin{center}
\includegraphics[width=0.9\textwidth]{{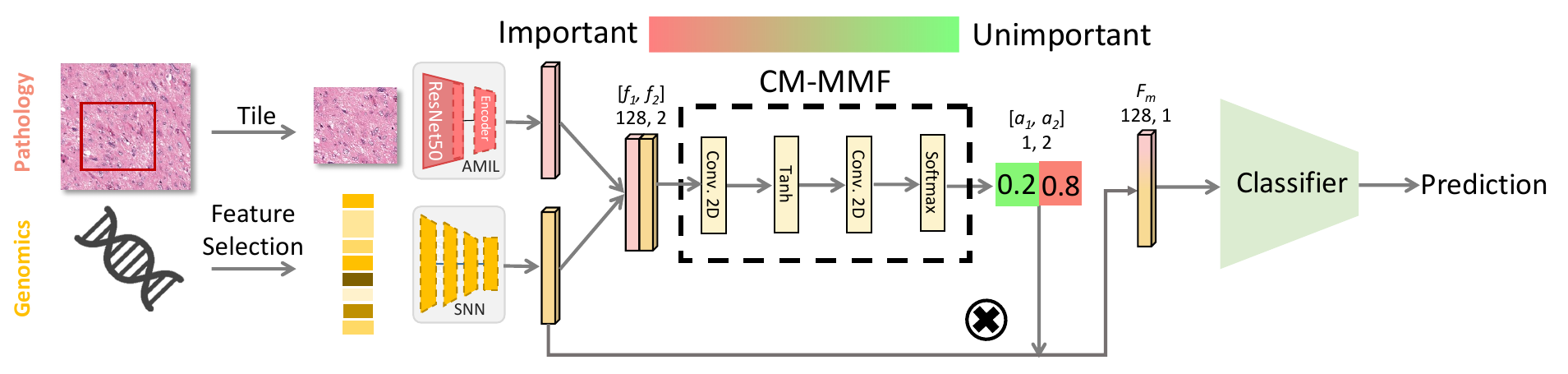}}
\end{center}
\caption{\textbf{Cross-modality attention-based Multimodal Fusion} This figure represents the pipeline to integrate the image and RNA-seq modalities.} 
\label{fig1:Problem}
\end{figure}

\section{Method}

\subsection{Unimodal Embedding}
The Attention Multiple Instance Learning (AMIL) module from the PORPOISE~\cite{chen2022pan} pipeline is 
utilized to transform each 1024-channel latent feature vector of image tile from a whole slide image (WSI) 
obtained through ImageNet pretrained ResNet-50 model,  into a 128-channel feature vector. The AMIL~\cite{ilse2018attention} module calculates an attention score for each tile based on its perceived relevance to patient-level prognostic prediction, enabling it to select pivotal tiles when aggregating the patient-level image feature representation. The Self-normalizing neural networks~\cite{klambauer2017self} (SNN) was employed to transform RNA-seq information into a 128-channel feature vector as omic feature representation. SNN was chosen due to its demonstrated superior performance in the unimodal setting with sequencing data.

\subsection{Cross-modality Attention-based Multimodal Fusion}
Upon receiving feature representations, which contain modality-specific knowledge from two modalities, normalization is undertaken as a preprocessing step prior to fusion. Inspired from one attention-based architecture for multi-scale disease classification~\cite{deng2022cross}, Cross-modality Attention-based Multimodal Fusion (CM-MMF) is introduced to weigh the significance of each modality for survival prediction. The CM-MMF comprises two fully convolutional layers with a kernel size of 1$\times$1, accompanied by a Tanh activation function. The kernel weights are shareable, promoting holistic learning of the importance of modality-specific knowledge through cross-modality relationships. The cross-modality attention ($a_m$) can be expressed as:

\begin{equation}
a_m = \frac{\exp{\mathbf{W}^\mathrm{T}\text{Tanh}(\mathbf{V}f_m^\mathrm{T})}}{\sum_{m=1}^M\exp{\mathbf{W}^\mathrm{T}\text{Tanh}(\mathbf{V}f_m^\mathrm{T})}}
\label{attention}
\end{equation}

\noindent Here, $\mathbf{W} \in\mathbb{R}^{L \times 1}$ and $\mathbf{V} \in\mathbb{R}^{L \times N}$ are trainable parameters in the CM-MMF, with $L$ representing the size of the unimodal embedding output $f_m$. Additionally, $N$ is the output channel of the first layer of CM-MMF, $\text{Tanh}(\cdot)$ denotes the tangent element-wise non-linear activation function, and $M$ signifies the number of modalities in the dataset.

\subsection{Multimodal Survival Prediction}
The cross-modality attention scores ($a_m$) are subsequently multiplied with corresponding modality  features to yield a unified cross-modality representation, as illustrated in Equation \ref{eq:cross-modality1}:

\begin{equation}
F_m = \sum_{m=1}^M a_mf_m
\label{eq:cross-modality1}
\end{equation}
For final prediction, a one-layer classifier adapted from PORPOISE~\cite{chen2022pan} is implemented to facilitate patient-wise survival prediction using cross-modality embedding ($F_m$).

\section{Data \& Experiments}
\subsection{Data}
In this study, we used data from patients which received atezolizumab plus carboplatin plus paclitaxel (also termed as ARM-A) for the first-line treatment of metastatic nonsquamous NSCLC from the IMpower 150 study~\cite{socinski2018atezolizumab}. This was part of a phase 3 clinical trial that evaluated the efficacy of adding targeted treatment to PD-L1 versus the current standard of care in NSCLC. To develop our multimodal framework, we worked with anonymized histopathology images (from 270 patients) alongside bulk RNA-seq data.


\textbf{Image Data} H\&E-stained WSI data were scanned at 20 $\times$ (0.5 micron/pixel) by an Aperio scanner. For the purpose of tile filtering, we first ran our in-house pretrained model with U-Net architecture to classify regions into tumor and stroma. 512 $\times$ 512 pixel tiles were then captured from those classified regions of WSIs, and then embedded into a 1024-channel feature vector using CLAM~\cite{lu2021data} feature extraction pipeline which uses ImageNet pretrained ResNet-50 model.

\textbf{RNA-seq Data} RNA-seq data, containing gene expression values along with Ensemble Gene IDs, were obtained from CID CIT Data MART. Considering the relatively small patients number, out of the 19K available genes, 154 genes were pre-selected through Elastic Net Cox model~\cite{wu2012elastic} fitting with 10 fold cross validation on the same data set.

\subsection{Loss Function}
Survival loss function~\cite{chen2022pan} is deployed for optimizing the outcome from the fusion architecture. The continuous timescale of overall patient survival time in days is partitioned into four non-overlapping bins. The negative log-likelihood (NLL) survival loss is used to supervise the training, using both censorship status and 4-bin interval labels as a classification task.

\subsection{Evaluation Metrics}
We assess the survival prediction results using the concordance index (c-index), where higher values indicate better performance. The c-index measures the proportion of all possible pairs of observations in which the model's predictions correctly order the actual survival times. All results from the baseline methods and our proposed method represent mean values of the c-index, calculated using 5-fold cross-validation on consistent data splits.

\subsection{Experiment Details}
To improve the training robustness, Gaussian noise was added to image features and RNA-seq features before loaded into the model. All of the models were trained over 55 epochs with a learning rate of 0.01 and a batch size of 1 using the ADAM optimizer. For each patient, all tiles from stroma and tumor regions were used for survival loss. Standardization was implemented for the RNA-seq modality, and normalization was deployed to rearrange the feature vectors between 0 and 1 for all modalities before implementing the fusion architecture. The remaining setting followed PORPOISE official pipeline~\footnote{https://github.com/mahmoodlab/PORPOISE}. 

\begin{table}[t]
    \centering
    \begin{adjustbox}{width=0.8\textwidth}
    \begin{tabular}{c|cc|c}
    \toprule
        Backbone & Fusion Design & Modality & C-index \\
    \hline
        AMIL~\cite{chen2022pan} & None & Unimodal - image & 0.5772 $\pm$ 0.0409 \\
        Custom-AMIL~\cite{chen2022pan} & None & Unimodal - image & 0.4906 $\pm$ 0.0430 \\
        SNN~\cite{klambauer2017self} & None & Unimodal - RNA-seq & 0.5885 $\pm$ 0.0259 \\
    \hline
        AMIL & Raw-concatenation* & Multimodal - image + RNA-seq & 0.6258 $\pm$ 0.0309 \\
        AMIL + SNN & Concatenation~\cite{mobadersany2018predicting} & Multimodal - image + RNA-seq & 0.6167 $\pm$ 0.0361 \\
        AMIL + SNN & Mean-vector~\cite{cheerla2019deep} & Multimodal - image + RNA-seq & 0.5087 $\pm$ 0.0411 \\
        AMIL + SNN & Bilinear~\cite{Chen_2021_ICCV} & Multimodal - image + RNA-seq & 0.5816 $\pm$ 0.0294 \\
        AMIL + SNN & Gated-attention~\cite{ilse2018attention} & Multimodal - image + RNA-seq & 0.6195 $\pm$ 0.0334\\
        AMIL + SNN & SurvPass~\cite{jaume2023modeling} & Multimodal - image + RNA-seq & 0.5427 $\pm$ 0.0271 \\
        AMIL + SNN & CM-MMF$\uparrow$ & Multimodal - image + RNA-seq & \textbf{0.6587} $\pm$ 0.0266 \\
    \bottomrule
    \end{tabular}
    \end{adjustbox}
    \caption{The C-index for Different Fusion Designs Across Various Modalities}
    \textbf{Raw-concatenation*}: RNA-seq features are directly concatenated with image features without passing through SNN.\\
    \textbf{CM-MMF$\uparrow$}: The proposed design which achieved better survival prediction in NSCLC.\\

    \label{tab:survivalprediction}
\end{table}

\section{Results}
AMIL~\cite{chen2022pan} and an custom-AMIL with adding more batch-normalization layers and ReLU activation were implemented as image unimodal encoders, while two deep learning networks, SNN~\cite{klambauer2017self} for RNA-seq data was deployed as RNA-seq unimodal encoders. AMIL and SNN were selected as the backbones for all fusion designs according to the better performance in unimodal training. 6 existing fusion approaches~\cite{mobadersany2018predicting,cheerla2019deep,Chen_2021_ICCV,ilse2018attention,jaume2023modeling} ranging from simple concatenation to attention-based architectures were deployed to compare the capabilities of the multimodal fusion with the proposed method. 

\subsection{Multimodal-fusion Results}
In table~\ref{tab:survivalprediction}, most of the fusion designs with multimodal learning achieved superior performance than unimodal learning, demonstrating the capability that infuse the modality-specific knowledge from different modalities. The proposed CM-MMF achieved better fusion performance supervised by survival loss, showcasing the functionality of multimodal fusion by considering cross-modality relationships. Directly utilizing the RNA-seq data in the fusion part (Raw-concatenation) yields stably superior performance, which can be attributed to the primary contribution from the RNA-seq modality.

\subsection{Ablation Study}
Inspired by ~\cite{yao2020whole} and ~\cite{ilse2018attention}, we explored various attention mechanism designs with different activation functions and evaluate those designs on the NSCLC dataset. We formed the CM-MMF into two strategies, differentiated by whether they shared the kernel weights while learning the embedding features from multiple modalities. As shown by the survival prediction performance in Table~\ref{tab:ablation1}, sharing the kernel weight in the CM-MMF with Tanh activation function achieved better performances with a higher mean value of c-index.




\begin{table}
\centering
\begin{tabular}{l|cc|c}
\toprule
Strategy & Modality Attention Layer & Activation Function & C-index\\
\midrule
1 & Non-sharing & ReLU & 0.6416 $\pm$ 0.0181\\
2 & Sharing & ReLU & 0.5735 $\pm$ 0.0114   \\
3 & Non-sharing & Tanh & 0.6329 $\pm$ 0.0356 \\
4* & Sharing & Tanh & \textbf{0.6587} $\pm$ 0.0266  \\
\bottomrule
\end{tabular}
\caption{Comparison of different fusion strategies in the multimodal paradigm}
\textbf{*}: The proposed design which achieved better survival prediction in NSCLC.\\
\label{tab:ablation1}
\end{table}

\subsection{Limitations}
In this study, we relied solely on one in-house dataset to gauge performance and observations. More data samples should be evaluated to gain a more generalized perspective on fusion performance. Meanwhile, other loss functions (e.g., cox loss, etc.) and performance metrics are expected to evaluate the supervisory prowess across different modalities more comprehensively. We primarily deployed only a few backbones for unimodal data, which might not be the optimal choices for unimodal representation. Implementing more embedding backbones could provide further insight into the capability of each unimodal learning method.

\section{New or Breakthrough Work to be Presented}
In this study, we propose a cross-modality attention-based multimodal fusion architecture (CM-MMF) to integrate knowledge from WSI and RNA-seq for enhanced lung cancer survival prediction. Within the fusion designs, our method assesses the importance of each modality for feature fusion, considering cross-modality relationships when amalgamating the multimodal features. This approach achieved the highest c-index 0.6587 in our experiment.

\section{Conclusion}
The proposed cross-modality attention-based multimodal fusion (CM-MMF) method outperformed other fusion designs and unimodal learning methods in this study. This underscores its capability to integrate modality-specific knowledge from various sources and highlights the functionality of multimodal fusion that takes cross-modality relationships into account. The attention scores from the fusion layer enable us to illustrate the significance of each modality for diagnosis. Meanwhile, the instance attention from AMIL can be used to indicate the contribution of each image tile. We will pursue detailed investigation in future work.

\section{ACKNOWLEDGMENTS} 
The work has not been submitted for publication or presentation elsewhere. We thank Jennifer Giltnane and Raghavan Venugopa for their support in providing the tissue image segmentation algorithm to facilitate tile extraction. We also thank the following individuals for their expertise and assistance throughout all aspects of this study and for their help leading to this manuscript: Ravi Kamble, Kamalakar Kodali, Qiangqiang Gu, Auranuch (Ney) Lorsakul, Xingwei Wang, and Marghoob Mohiyuddin.

\bibliography{main} 
\bibliographystyle{spiebib} 

\end{document}